# Controlling phase of microwaves with active graphene surfaces

Osman Balci, Nurbek Kakenov, Coskun Kocabas*

*Bilkent University Department of Physics, 06800 Ankara, Turkey*

*Corresponding author email: ckocabas@fen.bilkent.edu.tr*

In this letter, we report a method to control reflection phase of microwaves using electrically tunable graphene devices. The device consists of mutually gated large-area graphene layers placed at a quarter-wave distance from a metallic surface. This device structure yields electrically tunable resonance absorbance and step-like phase shift around the resonance frequency when the impedance of graphene matches with the free space impedance. Electrostatic control of charge density on graphene yields an ability to control both intensity (> 50 dB) and phase (~$\pi$) of the reflected electromagnetic waves with voltage. Furthermore, using the asymmetry of the doping at opposite polarity of the bias voltages, we showed bidirectional phase control with the applied voltage. We anticipate that our results will pave a new directions to control interaction of electromagnetic waves with matter for long wavelengths and could open a new avenue for microwave devices.

Controlling light-matter interaction has a central role in modern communication and information technologies. Electrical and magnetic properties of materials govern the physical mechanisms behind the interaction of light with matter. Ability to control material properties with electrical signal allows processing of information carried by light. Although sophisticated devices based on semiconductors and liquid crystals were developed for visible[1, 2] and infrared[3-5] part of electromagnetic spectrum, controlling light with longer wavelength (terahertz and microwaves) has been a long-standing challenge. The bottle-neck of this challenge is originated from the lack of an active material for long wavelengths. To overcome this challenge, metamaterials integrated with MEMS[6, 7], semiconductors[8, 9] and diodes[10-12] have been used to control amplitude and phase of terahertz and microwaves. Recent developments in the field of 2-dimentional (2D) materials provide a new perspective for active control of light in a very broad spectrum. 2D materials, especially graphene, provides a tunable platform for extremely the high mobility charge carriers which could yield tunable light-matter interaction. Tunable high mobility charge carriers together with the atomic thickness, give a unique combination to control light from visible to microwave frequencies. Graphene has been implemented for active terahertz devices using solid state[13-15] or electrolyte gating techniques[16-18]. Although these devices show large intensity modulation, the phase modulation have been negligible owing to the atomic thickness of graphene. Phase modulation can be achieved by tuning optical path or index of refraction of a medium. For atomically thin layers, however, even for large refractive index change, the phase modulation is extremely small due the short optical path ($t_g \approx 0.36 nm \ll \lambda$). Integrating these thin layers with a metamaterial structure is demonstrated as a good intensity and phase tuning device in IR and THz[19-22] region. Very recently, we have demonstrated active graphene devices operating in terahertz[23] and microwave frequencies[24]. We integrated electrically tunable graphene as a tunable absorbing layer and showed that these devices can operate as a perfect intensity modulator at their resonance frequencies. In addition to their intensity modulation ability, in this letter, we demonstrate another ability of these devices to control phase of electromagnetic waves in a broad frequency window near the resonance frequency.

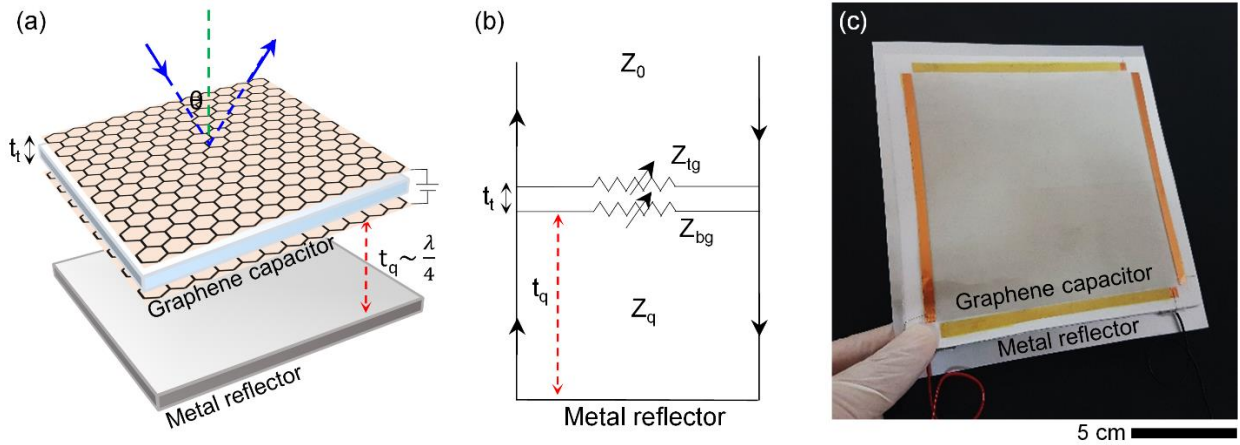

**Figure 1: Graphene based active Salisbury screen.** (a) Schematic layout of the active graphene device. Mutually gated graphene layers placed at a quarter-wave distance ($t_q$) from a metallic surface. (b) Transmission line representation of the active device. The graphene layers are represented with a tunable resistance (top graphene; $Z_{tg}$, bottom graphene; $Z_{tg}$) and the metallic surface is shown with a short circuit. (c) Photograph of the fabricated device. An ionic liquid electrolyte is sandwiched between two large-area CVD grown graphene on polymer substrates. The back side of the device is coated with an aluminum foil.

Our device is based on the idea of controlling phase shift due to a resonance. A step-like phase shift of π occurs in the response of a resonant system when frequency of external driving force is higher than the resonance frequency. Introducing a controlled damping in the resonator would enable to adjust the phase of its response. Here, we use electrostatically controlled graphene as a tunable absorber in a Salisbury screen thus to control phase of the reflected electromagnetic waves in a broad frequency band. Figure 1(a) shows the layout of the active graphene device. We fabricated a mutually gated graphene capacitor using two large area graphene and an ionic liquid electrolyte (DEME TFSI) sandwiched between them. Large area graphene was grown on ultra-smooth copper foils using chemical vapor deposition method at 1035°C. After the growth, we transferred graphene on microwave transparent polymer substrates using hot lamination method. A thin layer of optical tissue soaked with an ionic liquid electrolyte is sandwiched between the graphene layers. We placed the device on 5.3mm thick cardboard and coated the back side with an aluminum foil. Figure 1(c) shows the fabricated device attached with electrical wiring. This device operates as an active Salisbury screen. Transmission line model of the active device is shown in Figure 1(b) and explained in detail in supplemental material. Graphene layers are represented with two variable resistors ($Z_{tg}$, and $Z_{bg}$) and the metallic surface is shown as a short circuit at the end of the quarter-wave medium in the transmission line. The surface impedance of metals in microwave frequencies depends on the conductivity ($\sigma$) and skin depth ($\delta$) of the waves as $Z_q = 1/\delta\sigma$. Since the thickness of graphene ($t_g \cong 0.36 nm$) is much smaller than its skin depth in microwaves ($\delta \sim \mu m$), the surface impedance of the graphene layers can be represented with their sheet resistance ($R_s$) (supplemental material). Similarly the thickness of the electrolyte ($t_t \cong 100\mu m$) is much smaller than the wavelength, therefore the net input impedance of the graphene capacitor in the transmission line becomes half of the sheet resistance ($R_s/2$). Using this input impedance at a quarter wave distance ($t_q$) from a metallic surface which converts short circuit to open circuit, one can calculate the net total input impedance of the active Salisbury screen as (supplemental material);

$$Z_{total} = \frac{R_s \tan(k_q t_q)}{2\tan(k_q t_q) - i\frac{R_s}{Z_q}} \qquad \text{Eq. 1}$$

Where $Z_q = \sqrt{\mu_0/\varepsilon_q}$ is the wave impedance of the quarter-wave medium, $k_q = w\sqrt{\mu_0 \varepsilon_q}$ is the wave number and $\varepsilon_q$ is the dielectric constant of that medium. Using the expression for total input impedance, one can calculate the reflection coefficient of the device as $r = (Z_{total} - Z_0)/(Z_{total} + Z_0)$, the reflectance ($R = |r|^2$) and phase change ($\Delta\emptyset = arg(r)$). At the resonance the total input impedance of the device matches with the free space impedance ($Z_0 = 377\Omega$). Imaginary part of the impedance becomes zero and its real part becomes $Re\{Z_{total}\} = R_s/2$) at resonance. Therefore the reflection from the active Salisbury screen becomes zero and phase yields a $\pi$-phase-shifts.

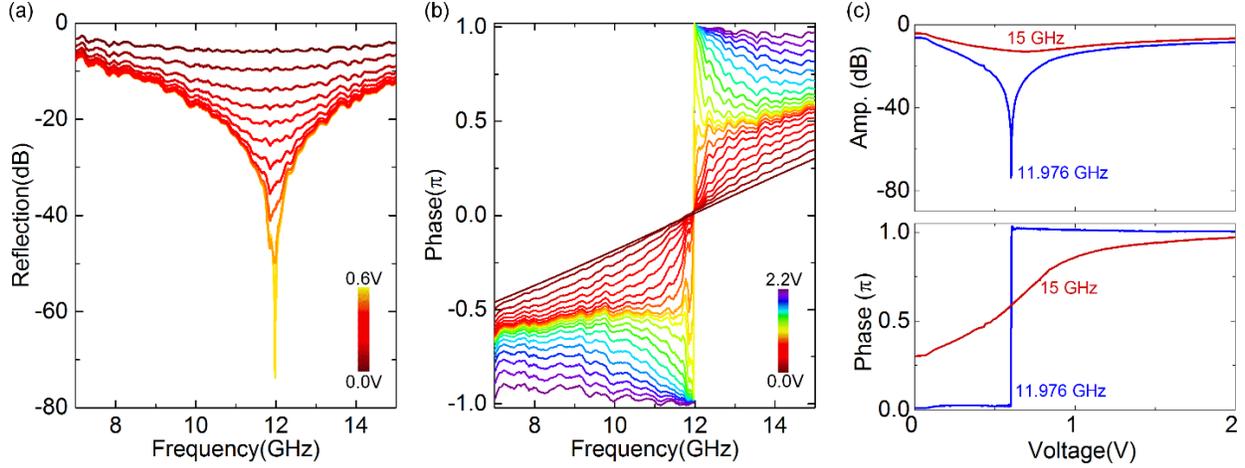

**Figure 2: Voltage controlled intensity and phase of reflected microwave:** (a) Spectrum of reflected electromagnetic waves from the graphene based Salisbury screen at bias voltages between 0 and -0.6 V. (b) Phase spectrum of the reflected electromagnetic waves at bias voltages between 0 and -2.2V. (c) Variation of the reflection amplitude and phase plotted against the bias voltage for on-resonance (11.976GHz) and off-resonance (15GHz).

Figure 2 summarizes microwave performance of the device. We measured the microwave reflectivity from the active device at different bias voltages. To measure reflectivity, we recorded two-port s-parameters using a network analyzer (Keysight E5063A) attached with two broadband standard gain horn antenna (ARRA Inc., BAY SHORE, serial:9625). Intensity of the recorded spectrum are given in Figure 2(a). As we gate graphene electrostatically, the reflectivity decreases and shows a strong resonance at 11.976 GHz. We observed the resonance at 0.6 V where half of the sheet resistance of graphene matches with free space impedance. More interestingly, phase of the reflected microwaves shows large variation with bias voltage (Figure 2(b)). We observed a step like phase modulation of π around the resonance frequency, $f_r$. The phase shift around a resonance is a universal behavior. However, in our configuration, tuning charge density on graphene enables us to manipulate the resonance condition and hence the phase shift near the resonance electronically. For frequencies less than the resonance frequency ($f < f_r$), the reflected microwaves show negative phase shift. However, for $f > f_r$, the phase shift has positive values. For these phase measurements, we took the plane of graphene capacitor as the reference plane.

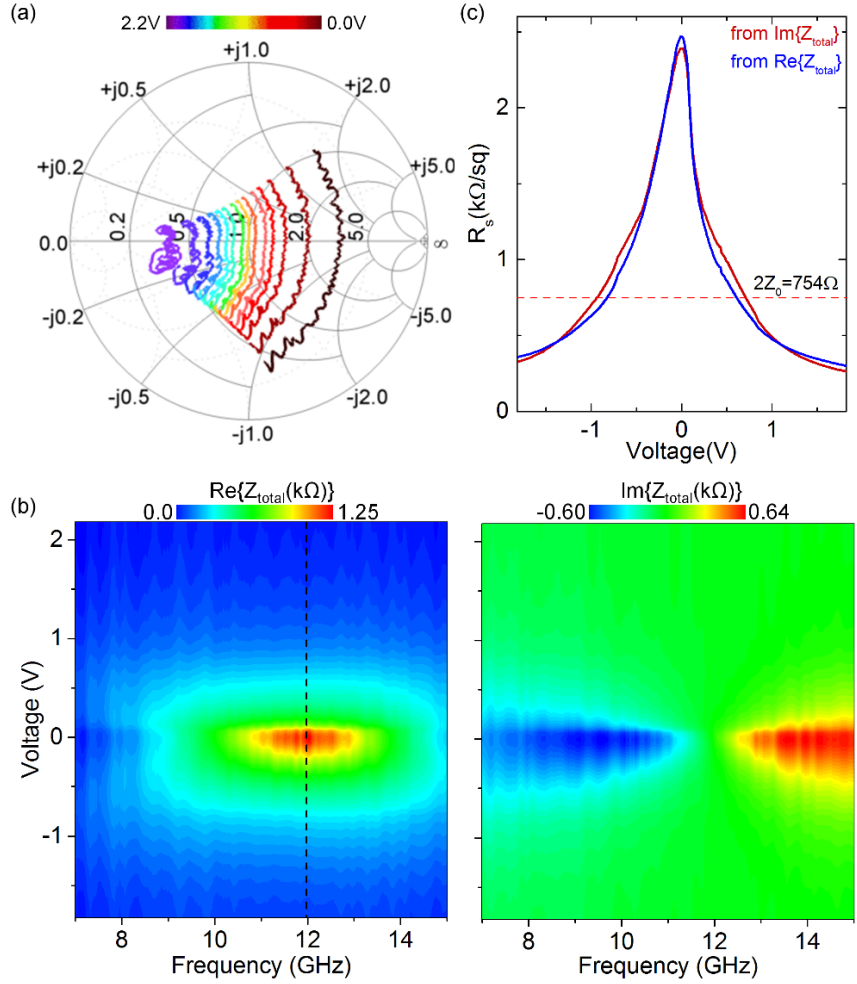

**Figure 3: Extracting sheet resistance of graphene from real and imaginary part of the input impedance:** (a) Smith chart showing the variation of total input impedance of our device with bias voltage. (b) Real and imaginary part of the total input impedance of the graphene based Salisbury screen as a function of both frequency and bias voltage. (c) Extracted sheet resistance of graphene from the total input impedance. Real part of the total input impedance at resonance frequency gives half of the sheet resistance. Imaginary part of the total impedance at its extremum gives one fourth of the sheet resistance of graphene at both left and right hand side of the resonance frequencies.

The Fresnel reflection coefficient (*r*) provides useful piece of information about the intrinsic properties of the graphene layer, such as sheet resistance and the characteristic impedance. In figure 3(a), we plot the measured Fresnel reflection coefficient of the active device using Smith chart representation (complex polar plot of reflection) for bias voltages ranging from 0 to 2.2V and frequencies between 7 and 15 GHz. At 0 V bias voltage (dark brown curve), both the reflection and corresponding impedance are high. The resonance condition is met at 0.6V when total impedance of the device matches the free space impedance. Increasing the bias voltage further to 2.2V decreases the impedance of the device therefore increases the reflection. More details about the microwave performance of the device is presented in supplemental material. Using the measured Fresnel reflection coefficient, we would like to calculate the complex impedance of the device ($Z_{total} = Z_0(1+r)/(1-r)$), which enables us to extract the sheet resistance of graphene. Eq. 1 provides the analytical form of the impedance of the device. At resonance frequency ($\lambda = 4t_q$), this expression has a real value which equals to the half of sheet resistance. 2-dimensional maps in Figure 3(b) shows the real and imaginary part of impedance of the device as a function of frequency and bias voltage. Real part yields a maxima of ~$1.25k\Omega$ at 0V at resonance frequency of 11.976 GHz. Variation of the impedance at resonance frequency (the vertical line in Figure 3(b)) gives an accurate measurement of the sheet resistance. Figure 3(c) shows the extracted values of the sheet resistance plotted against the bias voltage. It is also possible to use imaginary part of the total input impedance to extract sheet resistance of graphene. As shown in Figure 3(b), the imaginary part has two maxima at left and right hand side of the resonance frequency. At these extremum points, imaginary impedance yields quarter of the sheet resistance. We compare the extracted sheet resistance from real and imaginary part of the impedance in Figure 3(c). This method provides an accurate sheet resistance of large area graphene layers without a contribution from contact resistance.

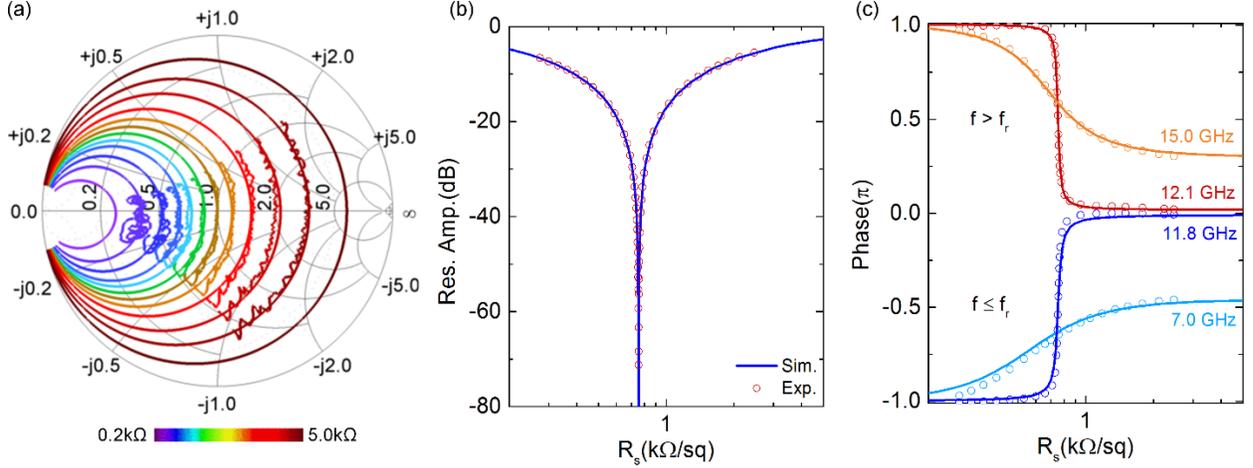

**Figure-4: Simulation of the active Salisbury screen.** (a) Simulated (straight lines) and measured (fluctuating lines) reflection coefficient (*r*) of graphene based Salisbury screen shown on a Smith chart. In the simulation, the frequency is changing between 1-23 GHz and the sheet resistance is changing between 0.2-5.0 kΩ. The frequency range for the experimental measurements is 7-15 GHz. (b) Comparison of experimental and calculated values of the reflection amplitude at resonance, plotted against the sheet resistance. (c) Variation of the phase at various frequencies plotted against the sheet resistance. Solid lines are achieved from the simulation.

To support our experimental observation, we developed a quantitative model for our device using transfer matrix formalism. Because of enormous mismatch between the thickness of graphene and wavelength of microwaves, Finite Element analysis is not convenient. To overcome this problem, we represented graphene as a conductive boundary. We calculated the Fresnel reflection coefficient from the boundary using the continuity of tangential components of electric ($\vec{n} \times (\vec{E}_2 - \vec{E}_1) = 0$) and magnetic ($\vec{n} \times (\vec{H}_2 - \vec{H}_1) = \vec{J}_s$) fields. The conductivity of the boundary is defined by frequency dependent conductivity of the graphene which includes inter-band and intra-band contributions as $\sigma(\omega) = \sigma_{inter}(\omega) + \sigma_{intra}(\omega)$. For microwave frequencies, the interband contribution is negligible therefore the total conductivity can be modeled with Drude conductivity which yields low frequency conductivity (ω<<1/τ) as $\sigma_{intra}(\omega) = \sigma_{DC}/1 - i\omega\tau \sim \sigma_{dc} = 1/R_s$. We calculated the Fresnel reflection coefficients as we change the sheet resistance of two graphene layers from 0.2kΩ to 5.0 kΩ in the supercapacitor geometry which is placed at a quarter-wave distance from a metallic surface. Figure 4 shows the comparison of the experimental and calculated results for the frequency range of 7-15 GHz. The Smith chart in Figure 4a demonstrates a nice agreement between experimental and calculated complex impedance values. When the sheet resistance of graphene is at $2Z_0=754\Omega$, resonance condition is met and the reflection becomes zero in both experiment and simulation. Decreasing the sheet resistance further increases reflection while total input impedance is decreasing. Using the extracted sheet resistance from Figure 3, the variation of the amplitude and the phase is plotted as a function of sheet resistance of graphene. We obtained excellent agreement between the model and experiments. Near the resonance there is a step-like sharp change in the phase while the change becomes softer at farther frequencies. Although the intensity shows large modulation in a narrow frequency window, the phase modulation expands significantly in a broader window. We believe this is a unique feature of the resonant device architecture.

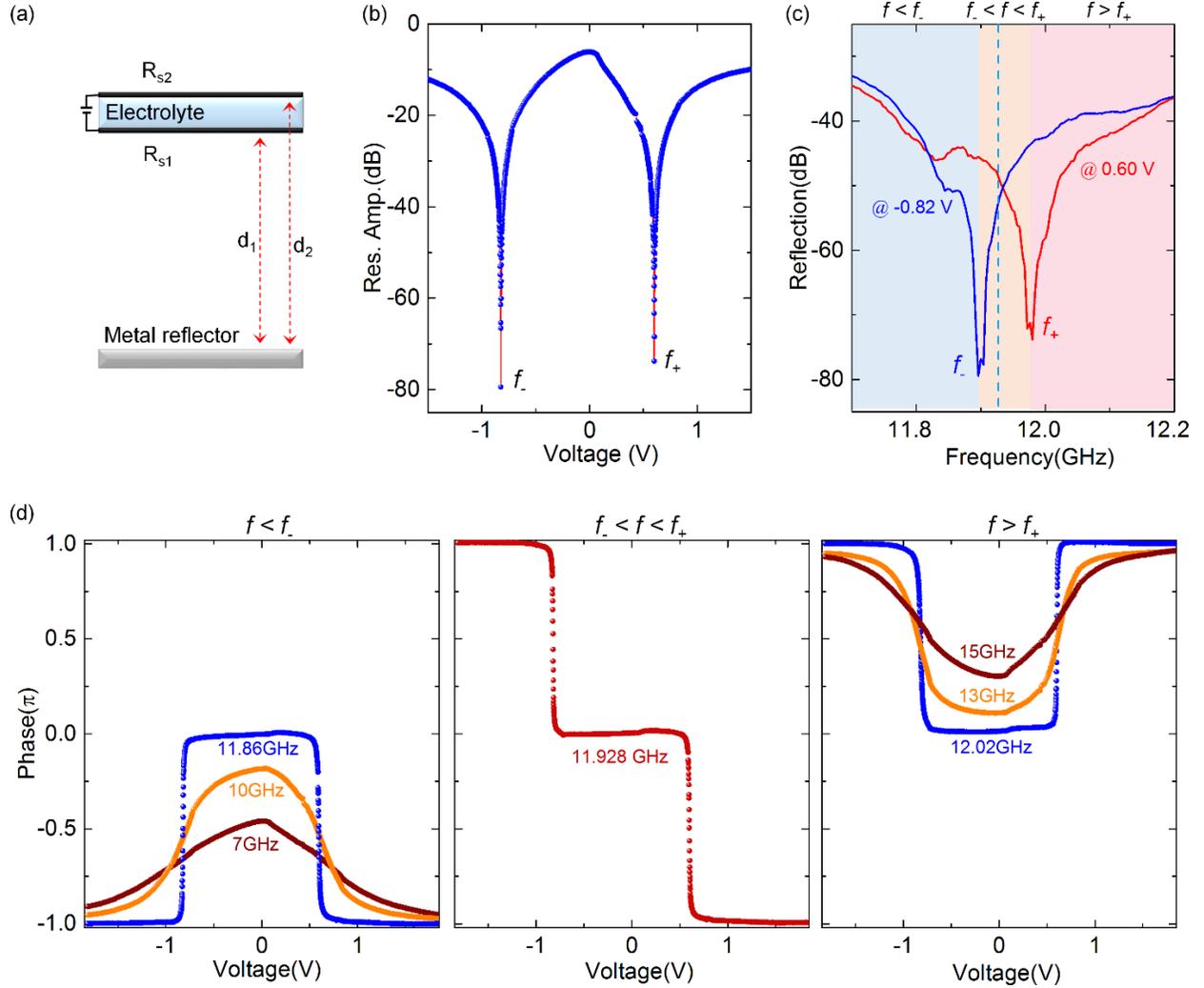

**Figure-5: Bidirectional phase control:** (a) Cross-sectional view of the device. Thickness of the electrolyte and the asymmetric doping between graphene layers yields two distinct resonance for positive and negative bias voltages. (b) Variation of the resonance amplitude with bias voltage, resonances at positive and negative bias voltages have similar amplitudes. (c) Spectrum of the resonance at positive and negative bias voltages. (d) Tunability of the phase with bias voltage for three different regions $f > f_+$, $f_+ < f < f_-$, $f > f_-$.

Our device provides an interesting capability for bidirectional phase control with the applied bias voltage owing to the device structure (Figure 5(a)). While we observed symmetric behavior for amplitude modulation for positive and negative bias voltages, we observed slightly different resonance frequencies for opposite polarity of bias voltages (Figure 5(b), (c)). Combination of slight electron-hole asymmetry (slightly larger hole doping due to ionic liquid gating, see Figure 3(c)) and the thickness of electrolyte medium, yields 0.1 GHz frequency shift between the resonances at positive and negative bias voltages ( Figure 5(c) ). Because of these distinct resonances ($f_+$, $f_-$), we observed three different behavior for the phase shifts with the bias voltage. In the first region $f < f_+$, we observed a negative phase shift independent of polarity of the bias voltage. For both electron and hole doping, we observed symmetric negative phase shift. Similarly, in the third region, $f > f_-$, we observed a positive phase shift for both polarization of the bias voltage. This behavior can be understood by the step like phase shift around the resonance. However, in the intermediate region, $f_+ < f < f_-$, where the frequency of the wave is between the two resonance frequencies, we observed asymmetric phase shift with the bias voltages. For this intermediate frequency region, when we apply positive voltage, the phase shift is positive, however, for negative voltages, the resonance frequency shift to $f_-$ which yields a negative phase shift. This is a unique advantage arising from the asymmetry of the doping and structure of our device. Although, the bidirectional phase control appears for a narrow frequency window, the width of the window can be enlarge by increasing the distance between the graphene layers.

As a conclusion, we report a s method to control phase of electromagnetic waves using active graphene based Salisbury screen. Our method relies on electrically-controlled resonance which yields a phase shift around the resonance. Self-gated graphene capacitor placed at a quarter-wave distance from a metallic reflector yields a resonant reflectivity which can be controlled by electrical gating. In this configuration graphene operates as a tunable absorbing layer. When the surface impedance of device matches with the free space impedance, we obtained the resonance condition yielding an absorption of 60dB and a phase shift of $\pi$. Using the measured complex reflection, we can extract the variation of sheet resistance of graphene layers with bias voltages. Furthermore, using asymmetry of the doping, we showed bidirectional phase control with the applied bias voltage. Our results provide a significant step in controlling light-matter interaction for long wavelengths and we anticipate that the capability of electrically controlled phase would open new directions for active microwave devices.

**Supplementary Material:** See supplementary material for the transmission line analysis and transfer matrix calculation for graphene based Salisbury screen in microwave frequencies. We modelled full structure of the device including two graphene layers and ionic liquid layer forming the graphene capacitors together with the spacing between the capacitor and the metal reflector. We also present the full microwave performance of our device for the bias voltages from 2.2V to -1.8V.

**Acknowledgment:** This work is supported by the European Research Council (ERC) Consolidator Grant ERC-682723 SmartGraphene.